\documentstyle[preprint,aps]{revtex}
\author{A. Gordon and J.~E.~Avron}
\address{ Department of Physics, Technion, 32000 Haifa, Israel}
\newcommand{\real}{{\rm I\kern-.2em R}}
\newcommand{\half}{\frac{1}{2}}
\newcommand{\zd}{\tiny{\frac{2}{3}}}

\title{Born-Oppenheimer Approximation near Level Crossing}

\begin{document}
\tightenlines
\draft
\maketitle

\begin{abstract}
We consider the Born-Oppenheimer problem near conical intersection
in two dimensions. For energies close to the crossing energy we
describe the wave function near an isotropic crossing and show
that it is related to generalized hypergeometric  functions
$_0F_3$. This function is to a conical intersection what the Airy
function is to a classical turning point. As an application we
calculate the anomalous Zeeman shift of vibrational levels near a
crossing.
\end{abstract}
%%%%%%%%%%%%%%%
\newpage
%%%%%%%%%%%%%%%%%%%%%%%%%
%\begin{multicols}{2}
\narrowtext
%%%%%%%%%%%%%%%%%%%%%%%%%%%%%%
%\section{Introduction}
%
{\bf Introduction.} The Born-Oppenheimer (BO) problem \cite{bo} is
concerned with the analysis of Schr\"odinger type operators where
the small electron to nucleon  mass ratio, plays the role of the
semiclassical parameter.
\cite{aventini-seiler,berry,lh,mead,mead2,mead_truhlar,jackiw,wilczek}.
The theory identifies distinct energy scales: The electronic scale
which, in atomic units, is of order one and the scale of nuclear
vibrations which is of order $(1/M)^{1/2}$ in these units. $M$ is
the nucleus to electron mass ratio. The identification of the
electrons as the fast degrees of freedom is central to the theory.

The clean splitting between fast and slow degrees of freedom fails
near eigenvalue crossing of the electronic Hamiltonian where there
is strong mixing between electronic and vibrational modes. This
lies at the boundary of the conventional BO theory. Since the
coupling between different electronic energy surfaces becomes
infinite near a crossing, the nuclear wave function does not
reduce to a solution of a scalar (second order) Schr\"odinger
equation.
%In \cite{mead2} the single surface nuclear wave near a crossing function is obtained,
% but the two coupled Schr\"odinger equations near it had never been analytically solved.
We describe the (double surface) nuclear wave function near an
isotropic conical crossing, for energies close to the energy of
the crossing.

The strong mixing of the electronic and nuclear degrees of freedom
near crossing leads an anomalous Zeeman effect. To describe the
anomaly recall that the Zeeman splitting in molecules is reduced
compared to the Zeeman splitting of atoms. It is convenient to
parameterize the reduction by a parameter $\gamma$ so that the
Zeeman splitting is of the form (and order) $ M^{-\gamma} B$ with
$B$ the external magnetic field. The low lying vibrational levels
have large reduction, $\gamma=1$. This is what one expects from
nuclei whose magnetic moments are by factor $M$ smaller than the
Bohr magneton. Levels {\em near the crossing energy} can have a
 small reduction expressed by the fact that $\gamma<1$. For
the isotropic situation we calculate $\gamma=1/6$, so that the
Zeeman shift is anomalously large, by a factor of about 2000, than
the normal Zeeman splitting of molecular levels. More precisely,
we find that the Zeeman splitting near isotropic crossing is
\begin{equation}\label{main2}
\Delta E(m)\approx \frac 1 {M^{1/6}}\frac{g(m)}{T_e}\,B.
\end{equation}
The sign $\approx$ means equality in the limit $M \to \infty$. $B$
is proportional to the magnetic field, with a coefficient
dependent only on the electronic wave functions at the crossing.
$m$, a half odd integer, is the azimuthal quantum number, $T_e$ is
an electronic time scale, see Eq.~(\ref{period}) below. $g(m)$ is
a universal dimensionless factor which is determined by the
nuclear wave function near the crossing, see Eq.~(\ref{gyro})
below. As we shall see $g(m)=-g(-m)$ and numerical estimates of
Eq.~(\ref{gyro}) give
\begin{equation}g\left(\half\right)= 0.961,\,\, g\left(\frac 3 2
\right)=0.543,\,\,g\left(\frac 5 2 \right)=0.396.\end{equation}
$g(m)$ is a molecular analog of the Land\'e g factor in atoms: So,
while Lang\'e g factor describes the Zeeman shift due to the
mixing of spin and orbital degrees of freedom, $g(m)$ does it for
the nuclear and electronic ones.

One can formulate the BO problem in the following way \cite{mead}:
\begin{equation}\label{Hbo}
H_{bo}= -\frac 1 {M}\Delta_x +H_e(x),
\end{equation}
where $H_e(x)$,  the electronic Hamiltonian, depends
parameterically on the nuclear coordinates $x$. When time reversal
is not broken, $H_e(x)$ is a real symmetric matrix \cite{spin}.
The Wigner von Neumann crossing rule \cite{wvn} says that $H_e(x)$
has generically a crossing point for two modes of vibrations,
$x\in\real^2$.
% in the case of integer spin and for five vibrational modes, $x\in\real^5$, for half integer spin.

Here we shall consider the simple scenario where $H_e(x)$ is a
$2\times 2$ matrix of  and $x\in\real^2$. This means that we shall
treat only the restriction of the electronic Hamiltonian to the
two dimensional subspace spanned by the two degenerate eigenstates
at the crossing point. We shall assume that $H_e(x)$ has a single
crossing point at $x=0$, and set the crossing energy at $0$.  We
further assume that the crossing is conic, that $H_e(x)$ is
isotropic about the origin, and that the $x$ dependence of
$H_e(x)$ is smooth near the origin. This is a common model
\cite{mead_truhlar}.

We first recall why the standard BO theory fails near a crossing.
When $H_e(x)$ is symmetric it can be diagonalized by an {\em
orthogonal} transformation $R(x)$. In the $2\times 2$ case, and
when $H_e(x)$ is non-degenerate, $R(x)$ is uniquely determined, up
to an overall sign, by requiring $\det R(x)=1$. Hence, away from
crossing, $H_{bo}$ is unitarily equivalent to
\begin{equation}\label{sHbo}
 -\frac 1 M\Big(\nabla_x+iA(x)\Big)^2
+\pmatrix{E_1(x)&0\cr 0&E_2(x)},
\end{equation}
where  $E_{1,2}(x)$ are the two eigenvalues of $H_e(x)$. The
vector potential $A(x)=-iR^T(x) \nabla_x R(x)$ is purely
off-diagonal. For linear crossing, $R(x)\to - R(x)$ as $x$
surrounds the origin \cite{lh}. This forces a $1/|x|$ singularity
of the vector potential for small $x$. Far from the origin, to
leading order in $1/M$, the two components of the wave function
decouple and are given by
\begin{eqnarray}\label{bo}
\Psi_{bo,j}(x)\approx\psi_{cl,j}(x) R(x)\pmatrix{\delta_{j,1}\cr
\delta_{j,2}}, \quad j=1,2,
\end{eqnarray}
where $\psi_{cl,j}$ is a semiclassical solution of the
Schr\"odinger equation with potential $E_j(x)$ in the cut plane
with antiperiodic boundary conditions \cite{lh,mead_truhlar}. This
decoupling holds provided $|x|>>M^{-1/3}$ \cite{prep}. We shall
henceforth denote $\epsilon=M^{-1/3}$, and refer to the region
$|x|>>\epsilon$ as "far from the crossing".  In contrast, the
divergence of the off diagonal part of $A$ near the crossing
prevents such a decoupling near the crossing. Our aim  is to
analyze the BO theory, to leading order in $\epsilon$, near the
crossing  of $H_e$. The reason one can still hope to say something
useful near the crossing is that the asymptotic form of $H_e(x)$
near the crossing, i.e. for $|x|<<1$, is universal \cite{mead2}
\begin{eqnarray}\label{can2}
H_e(x)= x_1 \sigma _1 + x_2 \sigma_3+O(x^2)
\end{eqnarray}
where $\sigma$ are the Pauli matrices.
% There can be a global coefficient in front of the r.h.s.~of (\ref{can2}) too, but this
% can be scaled out.
We shall refer to $|x|<<1$ as "close to the crossing". Notice that
our notions of close and far from the crossing have a nonempty
intersection. This enables us to match the solution close to the
crossing with the standard, decoupled, BO solution,
Eq.~(\ref{bo}).

{\bf The zero energy wave function close to the crossing.} We
study the wave functions near the crossing for energies close to
the crossing energy. Everything to be said from now on is true in
the limit of $M \to \infty$, to leading order in negative powers
of $M$.

We shall assume that zero is an eigenvalue of (\ref{Hbo}), since
there is always an eigenvalue that is close enough to zero
\cite{prep}. It turns out to be convenient first to
unitary-transform (\ref{can2}) with $e^{-i\frac \pi 4 \sigma_1}$.
This will replace $\sigma_3$ in (\ref{can2}) by $\sigma_2$. In
this representation, close to the crossing the zero energy wave
function satisfies approximately the differential equation
\begin{equation}\label{scaled2}
\left\{ -\nabla_\xi^2 + \xi_1 \sigma _1 + \xi_2
\sigma_2\right\}\Psi(\xi_1,\xi_2) = 0
\end{equation}
$\Psi$ stands for a two component column matrix and
$\xi_i=M^{1/3}\,x_i$ is a scaled variable.

The operator $J_3=L_3+\half\sigma_3=-i \xi_1\partial_{\xi_2}+i
\xi_2\partial_{\xi_1}+\half\sigma_3$ commutes with the operator on
the l.h.s.~of (\ref{scaled2}). It does not have the meaning of
total angular momentum, since the Pauli matrices do not represent
spin. We thus consider solutions of the differential equation
(\ref{scaled2}) which are eigenfunctions of $J_3$ with an
eigenvalue $m$, namely:
\begin{eqnarray}\label{radial}
\Psi(\xi;m)=e^{im\theta}\, \pmatrix
  {\varphi_1(\rho;m)e^{-i\theta/2}  \cr
 \varphi_2(\rho;m)e^{i\theta/2}}
\end{eqnarray}
$\rho, \theta$ are the polar coordinates related to $\xi$. $m$
must be {\it half odd integer}, for the wave function to be single
valued. Separating variables, the radial equation obtained from
(\ref{scaled2}), in the $m$-th sector, takes the form:
\begin{eqnarray}\label{scaled}
   \left\{  -\frac{d^2}{d\rho^2}-\frac{1}{\rho}\frac{d}{d\rho}+
  \frac{m^2+\frac 1 4}{\rho^2}
+\pmatrix{ -\frac{m}{\rho^2} & \rho \cr \rho &
\frac{m}{\rho^2}}\right\}\pmatrix{\varphi_1 \cr \varphi_2}=0
\end{eqnarray}
%Our aim in this section is to solve, exactly, Eq.~(\ref{scaled}).
Of the four linearly independent solutions of Eq.~(\ref{scaled}),
basically due to boundary conditions, only one linear combination
is fit to represent a wave function.  We denote it by ${\cal
F}_c(m,\rho)$ and its components by $\varphi_{1 c}(m;\rho)$ and
$\varphi_{2 c}(m;\rho)$.
%It is finite at the crossing point, and it propagates into the region $\rho
% >>1$, far from the crossing, where it matches the
%BO wave function $\Psi_{bo,1}$ of Eq.~(\ref{bo}).
${\cal F}_c(m,\rho)$ is to a crossing point what the Airy function
is to a classical turning point \cite{landau,powell}: It
interpolates between the near region where the wave function is
intrinsically a two component spinor, and the far region where the
wave function is highly oscillatory and given by Eq.~(\ref{bo}).
${\cal F}_c(m,\rho)$ has a closed expression in terms of the
generalized hypergeometric functions of the kind $_0F_3$. It has
the asymptotic (for $\rho>>1$) form
\begin{equation}\label{cos}
{\cal F}_c(m,\rho) \approx \frac{1}{\rho^{3/4}}\,\cos
\left(\zd\rho^{3/2}-\pi\left(\frac m 3 + \frac 1
4\right)\right)\pmatrix{1\cr -1},
\end{equation}
%We shall now describe the main idea of solving (\ref{scaled}).
Solving (\ref{scaled}) asymptotically near the origin
gives\cite{boyce}
\begin{equation}\label{origin}
\pmatrix{\varphi_1(\rho;m)\cr \varphi_2(\rho;m)} \sim \pmatrix{a_+
\rho^{m-1/2}+a_-\rho^{-m+1/2} \cr b_+
\rho^{m+1/2}+b_-\rho^{-m-1/2}}
\end{equation}
Solving (\ref{scaled}) asymptotically at infinity we obtain
\begin{eqnarray}\label{infinity}
\pmatrix{\varphi_1(\rho;m)\cr \varphi_2(\rho;m)}
\sim\frac{1}{\rho^{3/4}}\left\{\left(A_+e^z+A_-e^{-z}\right)\pmatrix{1\cr
1} +C\cos \left(z+\phi\right)\pmatrix{1\cr -1}\right\},\quad
z=\zd\rho^{3/2}
 \end{eqnarray}
The four dimensional family of solutions can be parameterized  by
either $a_+, a_-, b_+, b_-$ or $A_+, A_-, C, \phi$. Requiring the
solution to be bounded at the origin and at infinity means (for
positive $m$) that $A_+=0, a_-=0, b_-=0$. Imposing three
homogeneous conditions on a four dimensional linear space leaves
us with one dimensional subspace, i.\,e.~a certain function times
an arbitrary  constant. This is the celebrated ${\cal
F}_c(\rho,m)$.

While the reason we require ${\cal F}_c(\rho,m)$ to be bounded at
the origin is obvious, the reason we require it to be bounded at
infinity is a bit subtle, since (\ref{scaled}) is meaningful only
close to the crossing. However, "close to the crossing",
$\rho<<M^{1/3}$, extends farther and farther in terms of $\rho$ as
$M$ gets large. $A_+$ at (\ref{infinity}) should be exponentially
small, and can set to zero to leading order.

The solutions which are regular at the origin can be obtained from
the fourth order differential equations, one for each component
$\varphi_j$, obtained from (\ref{scaled}). These are related to
the differential equation that defines the generalized
hypergeometric functions $_0F_3$.
%Namely, with $\zeta=\rho^6/6^4$, and ${\rm
%D}=\zeta\frac{d}{d\zeta}$:
%\begin{eqnarray}\label{4order}
%\left\{{\rm D}({\rm D}-\frac{1}{2}+\frac{(-1)^j}{6}) ({\rm
%D}+\frac{m}{3}-\frac{1}{2}) ({\rm
%D}+\frac{m}{3}+\frac{(-1)^j}{6})-\zeta\right\}
%\zeta^{\frac{-m-(-1)^j/2}{6}}\varphi_j(\zeta)=0.
%\end{eqnarray}
The two linearly independent solutions that are regular at the
origin are \cite{prep}:
\begin{eqnarray}\label{hyper_solutions1}
{\cal F}_1(\rho;m)&=& \pmatrix{
  \rho^{m-\frac{1}{2}}{ _0F_3}(
  ;\frac{1}{3},\frac{1}{2}+\frac{m}{3},\frac{5}{6}+
  \frac{m}{3};\frac{\rho^6}{6^4})\cr \frac{\rho^{m+\frac{5}{2}}}{6+4m}{ _0F_3}(
  ;\frac{4}{3},\frac{3}{2}+\frac{m}{3},\frac{5}{6}+
  \frac{m}{3};\frac{\rho^6}{6^4})};\cr \cr
  {\cal F}_2(\rho;m)&=&\pmatrix
  {
  \frac{\rho^{m+\frac{7}{2}}}{12+8m}{ _0F_3}(
  ;\frac{5}{3},\frac{3}{2}+\frac{m}{3},\frac{7}{6}+
  \frac{m}{3};\frac{\rho^6}{6^4})\cr \rho^{m+\frac{1}{2}}{ _0F_3}(
  ;\frac{2}{3},\frac{1}{2}+\frac{m}{3},\frac{7}{6}+
  \frac{m}{3};\frac{\rho^6}{6^4})},
\end{eqnarray}
where $_0F_3(;a,b,c;\rho)$ are generalized hypergeometric
functions \cite{handbook,edm}. The linear combination
\begin{equation}\label{F}
{\cal F}_c(\rho;m)= A_1(m){\cal F}_1(\rho;m)+ A_2(m){\cal
F}_2(\rho;m)
\end{equation}
is bounded at infinity provided \cite{prep}:
\begin{equation}\label{A}
A_j(m)=\frac{-(-1)^j
2\pi^{3/2}6^{\frac{(-1)^j-2m}{3}}}{3\Gamma(\frac 1 2 +\frac m
3)\Gamma(\frac 1 2
-\frac{(-1)^j}{6})\Gamma(1+\frac{2m-(-1)^j}{6})}
\end{equation}
We have set the global coefficient in front of (\ref{F}) such that
(\ref{cos}) will be true. Formulae
(\ref{hyper_solutions1}-\ref{A}) are correct only for positive
$m$-s. Their negative counterparts can be obtained by
interchanging the lower and upper components, as can be seen from
(\ref{scaled}).

{\bf The anomalous Zeeman shift.} Let us now turn to the Zeeman
shift. The magnetic field breaks time reversal symmetry, which
means that it adds an the imaginary term to $H_e(x)$, in the
representation where $H_e(x)$ is real. Generically the magnetic
field will remove the conical intersection of $H_e(x)$ at $x=0$
and create a gap between the two energy sheets. The gap will be
proportional to the magnetic field in atomic units, and the
coefficient will be in general of order one. We can therefore
introduce the magnetic field into our model by adding the term
$B\sigma_2$ to $H_e(x)$ at (\ref{can2}). There is no harm in
taking $B$ to be independent of $x$, since only the value of $B$
in the origin will be of importance to us. $B$ is therefore a
constant, proportional to the magnetic field in atomic units. We
shall not minimally couple the magnetic field directly to the
vibrations for this turns out to be a weaker effect, of order
$M^{-1}$, while the shift mediated by the electrons, in the
rotationally invariant case, is $M^{-1/6}$, as we shall see.

Let us consider the case where the rest of $H_e(x)$, namely the
$O(x^2)$ part in (\ref{can2}), does not break the full rotational
symmetry of its linear part, so that $m$ is still a good quantum
number.

 In the rotational invariant case, the model has, for $B=0$,
a two-fold degeneracy: The states $m$ and $-m$ are degenerate. The
magnetic field $B$ breaks this degeneracy. The splitting is twice
the Zeeman shift in the energy for the two states $\pm m$ move in
opposite directions.

Equipped with approximants to the wave function near and far from
the crossing, with a degenerate perturbation theory one can
calculate, to leading order in $1/M$, the Zeeman splitting and
obtain (\ref{main2}). We describe this calculation in details
elsewhere \cite{prep}. Here we would only like to sketch the
derivation of the power in (\ref{main2}).

Far from the crossing, one can neglect the vector potential in
(\ref{sHbo}). The WKB approximant to the radial part of
$\psi_{cl,2}$ is
\begin{equation}\label{psi_cl}
\psi_{cl,2}(r) \approx \frac{N}{r^{3/4}}\cos(\sqrt{M}\frac 2 3
r^{3/2}+\phi),
\end{equation}
where we have employed the linearity of the energy surfaces near
the crossing. It is a general property of the WKB approximation,
that the normalization coefficient $N$ is independent of $M$, to
leading order in $1/M$.
% It is easily proven for scalar Schr\"odinger equations, and is inherited by "spinor" equations.
From (\ref{cos}) and (\ref{psi_cl}) one sees that the
interpolation of the BO radial wave function towards the crossing
is
\begin{equation}\label{interp}
\Psi(r;m)\approx NM^{1/4}{\cal F}_c(m;M^{1/3}r).
\end{equation}
$B$ removes the degeneracy of the electronic levels at $x=0$. The
gap created there due to $B$ is equal to $2B$. Intuitively, the
Zeeman shift of a vibrational level will be proportional to the
amount of probability density in the vicinity of the crossing
times $B$. By ``vicinity" we mean a neighbourhood  of order
$M^{-1/3}$ of the origin, the area of which is of order
$M^{-2/3}$. The density associated with the wave function is large
there and by (\ref{interp}) is proportional to $M^{1/2}$. Hence
the weight is proportional to $(M)^{-1/6}$ which gives the power
of the Zeeman splitting in Eq.~(\ref{main2}). The coefficient of
proportion will include an integral over the components of $\cal
F$, \cite{prep}, which gives $g(m)$
\begin{equation}\label{gyro}
g(m)=\int_0^\infty \rho\, d\rho\, (\varphi_{1c}^2(\rho;m)-
\varphi_{2c}^2(\rho;m)).
\end{equation}
N gives the factor $1/T_e$, where $T_e$ is
\begin{equation}\label{period}
 T_e =\int dr
 \frac{1}{\sqrt{-E_1(r)}},
\end{equation}
The integration is carried out between the two turning points of
$E_1$, the negative energy sheet  (\ref{sHbo}). Eq.~(\ref{period})
is proportional to the time in takes a particle with a unit
(electronic) mass to travel  classically across the potential. It
is independent of the nuclear mass, and $T_e^{-1}$ has the order
of magnitude of electronic energies. From the invariance under $T$
of $H_{bo}(B=0)$ it follows that $g(m)=-g(-m)$.

One  motivation for this work was an attempt to gain some
understanding of the different status of crossing in theory and
experiment. Theory puts crossing and avoided crossing in distinct
baskets: conic crossings come with fractional azimuthal quantum
numbers while avoided crossings come with integral quantum
numbers. In contrast, measurements of molecular spectra normally
can not tell a crossing from near avoided crossing. Only with {\em
precision} measurements \cite{bush} and {\em precise} quantum
mechanical calculations \cite{kendrick} can one tell when
molecular spectra favor an interpretation in terms of crossing and
half integral quantum numbers or avoided crossing with integral
quantum numbers. Zeeman splitting appears to be a useful tool to
study crossing. The anomaly of crossing is characterized by a
fractional power $\gamma$ of the molecular reduction of the Zeeman
splitting, $M^{-\gamma}$. A system of choice for studying crossing
is molecular trimers \cite{bush,kendrick}. Since trimers are not
rotationally invariant Eq.~(\ref{main2}) does not apply and we can
not conclude that near crossing $\gamma=1/6$ for trimers. However,
it is natural to expect that the qualitative features of our
results carry over also to the non-isotropic case where crossing
will manifest itself by an anomalously large Zeeman splitting and
a {\em fractional} $\gamma$. It is an interesting challenge to
calculate or measure the value of $\gamma$ for (other) molecular
crossings, and trimers in particular.

 \section*{Acknowledgments}
We thank M.V. Berry for encouraging us to look for a special
function that characterizes the crossing and E. Berg, A. Elgart
and L. Sadun for helpful suggestions. We thank C.~A.~Mead for
useful comments. This research was supported in part by the Israel
Science Foundation, the Fund for Promotion of Research at the
Technion and the DFG.

%\end{multicols}
\begin{figure}[htb] \center {\input epsf \epsfbox{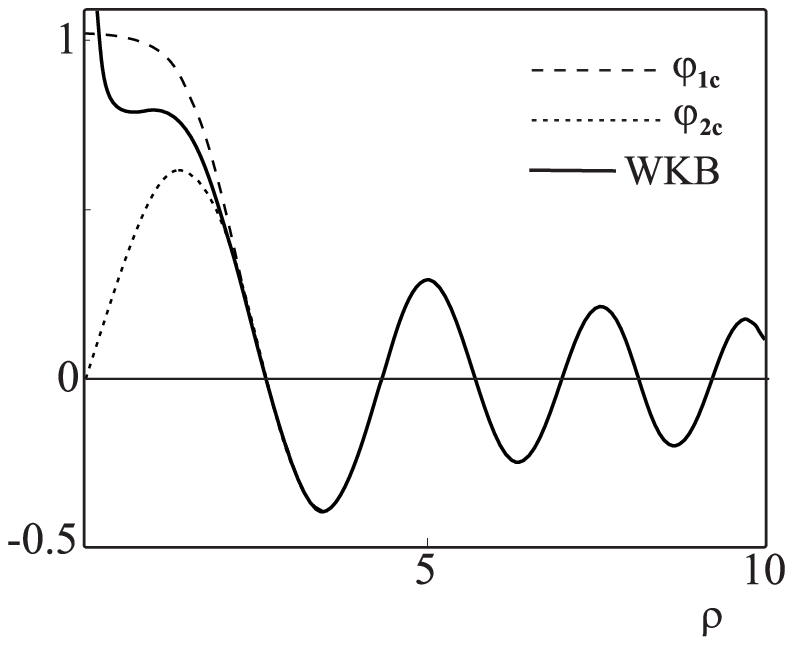}}
\caption{The components of ${\cal F}_c(m;\rho)$ for $m=\frac 1 2$
and their asymptotic form given by (\ref{cos}), which is also the
BO wave function approximated by WKB.} \label{f1}
\end{figure}

\end {document}